%% file: main.tex
\documentclass[10pt, conference]{IEEEtran}
\IEEEoverridecommandlockouts
\IEEEpubid{\makebox[\columnwidth]{978-1-6654-3924-4/21/\$31.00~\copyright2021 IEEE \hfill} \hspace{\columnsep}\makebox[\columnwidth]{ }}
\usepackage{cite}
\usepackage{amsmath,amssymb,amsfonts}
\usepackage{graphicx}
\usepackage{textcomp}
\usepackage{xcolor}
\usepackage[english]{babel}
\usepackage{tikz}
\usepackage{amsmath}
\usepackage{filecontents}
\usepackage{threeparttable}
\usepackage{hyperref}
\usepackage{graphicx}
\usepackage{caption}
\usepackage{subcaption}
\usepackage{verbatim}
\usepackage{multicol}
\usepackage{booktabs}
\usepackage{filecontents}
\usepackage{float}
\usepackage{listings}
\usepackage{cite}
\usepackage[ruled,vlined,linesnumbered]{algorithm2e}
    
    \SetCommentSty{mycommfont}
    
\restylefloat{table}
\usepackage{longtable}
\usepackage[T1]{fontenc}
\usepackage{amsmath}
\usepackage{algpseudocode}
\usepackage{xcolor}

\usepackage{enumerate}
\usepackage[shortlabels,inline]{enumitem}

\definecolor{codegreen}{rgb}{0,0.6,0}
\definecolor{codegray}{rgb}{0.5,0.5,0.5}
\definecolor{codepurple}{rgb}{0.58,0,0.82}
\definecolor{backcolour}{rgb}{0.95,0.95,0.92}

\lstdefinestyle{jackcodestyle}{
    language=go,
    numbers=left,
    stepnumber=1,
    numbersep=5pt,
    tabsize=2,
    showspaces=false,
    showstringspaces=false,
    basicstyle=\ttfamily\footnotesize,
    breakatwhitespace=false,         
    breaklines=true,                 
    captionpos=b,                    
    keepspaces=true,                 
    numbers=left,                    
    numbersep=5pt,                  
    showspaces=false,                
    showstringspaces=false,
    showtabs=false,
    backgroundcolor=\color{backcolour},   
    commentstyle=\color{codegreen},
    keywordstyle=\color{magenta},
    numberstyle=\tiny\color{codegray},
    stringstyle=\color{codepurple},
}

\def\BibTeX{{\rm B\kern-.05em{\sc i\kern-.025em b}\kern-.08em
    T\kern-.1667em\lower.7ex\hbox{E}\kern-.125emX}}

\algnewcommand\algorithmicforeach{\textbf{for each}}
\algdef{S}[FOR]{ForEach}[1]{\algorithmicforeach\ #1\ \algorithmicdo}

\newcommand{\coin}[1]{\texttt{#1}}
\newcommand{\xrpaddr}[2][\tiny]{{#1\href{https://xrpscan.com/account/#2}{\texttt{#2}}}}

\addto\extrasenglish{%
}	
\begin{document}

\title{\Large \bf{\textsc{Jack the Rippler}}: Arbitrage on the Decentralized Exchange of the XRP Ledger
}

\author{\IEEEauthorblockN{
Gaspard Peduzzi}
\IEEEauthorblockA{\textit{APWine}\\
Lausanne, Switzerland \\
gaspard@apwine.fi}
\and
\IEEEauthorblockN{
Jason James}
\IEEEauthorblockA{\textit{UCL} \\
London, UK \\
jason.james.19@ucl.ac.uk}
\and
\IEEEauthorblockN{
Jiahua Xu}
\IEEEauthorblockA{\textit{UCL Centre for Blockchain Technologies} \\
London, UK \\
jiahua.xu@ucl.ac.uk}
}

\maketitle

\IEEEpubidadjcol

\begin{abstract}
The XRP Ledger (XRPL) is a peer-to-peer cryptographic ledger. It features a decentralized exchange (DEX) where network participants can issue and trade user-defined digital assets and currencies.
We present {\sc Jack the Rippler}, a bot that identifies and exploits arbitrage opportunities on the XRPL DEX. We describe {\sc Jack}'s arbitrage process and discuss risks associated with using arbitrage bots. 
\end{abstract}


\input{content}

\section*{Acknowledgment}

This paper is based on {\tt SpringBlock Labs}' entry in the 2019 Block-Sprint Hackathon. We thank teammates Hugo Roussel, Henry Declety and Matthieu Baud for their contribution to the early development of the software.\footnote{\url{https://github.com/GaspardPeduzzi/spring_block}}

\bibliographystyle{IEEETranS}
\bibliography{references}

\end{document}

%% file: content.tex
\section{Background}

\subsection{Cross-currency arbitrage}
Arbitrage is the simultaneous purchase and sale of an asset to profit from an imbalance in price. It is a trade that profits by exploiting the price differences of identical or similar financial instruments on different markets or in different forms. In the conventional currency market, arbitrage is difficult to execute due to various market frictions such as excessive order book spread, high transaction fees, capital controls, etc. On the XRP Ledger (XRPL)'s decentralized exchange (DEX), users are able to conduct cross-currency payments with little friction thanks to its high speed and low transaction fees.

In this study, we illustrate the existing arbitrage behavior on the DEX of XRPL, and present \textsc{Jack the Rippler}, a bot that takes arbitrage opportunities on XRPL.

\subsection{Key XRPL components}

\subsubsection{Transaction types}
Multiple types of transactions can be conducted on XRPL:
\paragraph*{{\tt OfferCreate}} an order for a given currency pair.
\paragraph*{{\tt OfferCancel}} cancellation of an existing order.
\paragraph*{{\tt Payment}} a peer-to-peer payment or a fulfillment of an existing order.
\paragraph*{{\tt TrustSet}} setting a trustline between two accounts. 

\subsubsection{Flags}

XRPL transactions apply different flags that dictate a transaction's behavior: 
\paragraph*{{\tt tfPartialPayment}} a ``best flow effort" flag that allows partial payment and prevents the transaction being cancelled in case of insufficient liquidity on the path or insufficient balance in the recipient trustline.
\paragraph*{{\tt tfNoDirectRipple}} specifying the exact path (ordered list of offers) to be used when consuming the transaction. This is intended to force the transaction to take the arbitrage opportunity.

\subsubsection{Trustlines}

The success of a {\tt Payment} transaction necessitates the existence of a trustline established by the asset receiver towards the asset issuer. A trustline can either be a direct linkage between the receiver and the issuer, or, can be an indirect path if the flag {\tt tfNoDirectRipple} is not activated.
Absent a trustline, a payment transaction fails with {\tt tecPATH\_DRY} error.
\coin{XRP} is the only asset on XRPL that does not require a trustline for successful transactions.


\subsubsection{Assets}

With the exception of the native currency \coin{XRP}, a digital asset on XRPL exists in the form of an IOU (``I owe you'') \cite{Perez2020e}, and is primarily characterized by
\begin{enumerate*}[label={(\roman*)}]
\item a currency code (i.e. ticker), e.g. \coin{BTC}, \coin{USD}, \coin{CNY}; and
\item the introduction of further criteria or constraints.
\end{enumerate*}

Anyone can act as a gateway on the XRP ledger and issue their own IOU tokens.
An XRPL account must create a trustline to a gateway through a {\tt TrustSet} transaction to indicate willingness to hold a specific currency issued by the gateway.
Ripple provides a list of featured issuers, based on their business practices, volume, and other measures. Yet there is no formal process for ascertaining whether or not a gateway can be trusted, and whether their issued IOU tokens are ultimately redeemable.

\section{Arbitrage on XRPL}

\begin{figure}[tb]
\begin{center}
\includegraphics[width=\columnwidth]{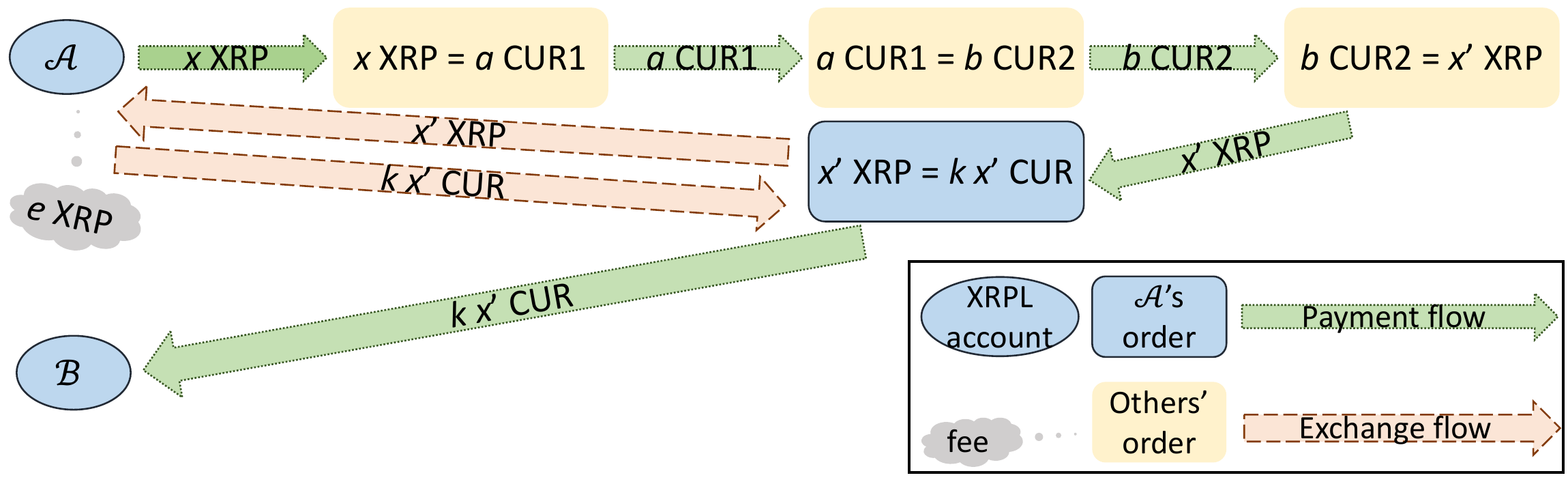}
\caption{Stylized arbitrage on XRPL}
\label{fig:speccase}
\end{center}
\end{figure}

XRPL does not allow a cross-currency {\tt Payment} cycle from \coin{XRP} to \coin{XRP}. Therefore, two transactions must be performed to complete an \coin{XRP} to \coin{XRP} arbitrage cycle. \autoref{fig:speccase} illustrates a stylized \coin{XRP} to \coin{XRP}  arbitrage scheme on XRPL, where accounts $\mathcal{A}$ and $\mathcal{B}$ are controlled by the same entity. Specifically, $\mathcal{A}$ first performs an {\tt OfferCreate} transaction with the intention to buy \coin{XRP} with self-issued currency \coin{CUR} that only $\mathcal{B}$ has the necessary trustline to accept. $\mathcal{A}$ then sends a cross-currency {\tt Payment} to $\mathcal{B}$ where $\mathcal{A}$ pays \coin{XRP} and $\mathcal{B}$ receives \coin{CUR}, triggering the fulfillment of $\mathcal{A}$'s outstanding order. \autoref{fig:speccase} shows that the aggregate gain of $\mathcal{A}$ and $\mathcal{B}$ equals $x'-x-e$. Naturally, the process is only profitable if this number is positive.
Historically, this scheme is found to be heavily used by \xrpaddr{rhAenHTxnAAxpMsrWorCcgfdzjyg6zsXRP} and \xrpaddr{rsdMbYxHmYswHCg1V6vBsnxmHuCjpn6SC4}.

\subsection{Arbitrage process with {\sc Jack the Rippler}}

\begin{figure}[tb]
\begin{center}
\includegraphics[width=\linewidth]{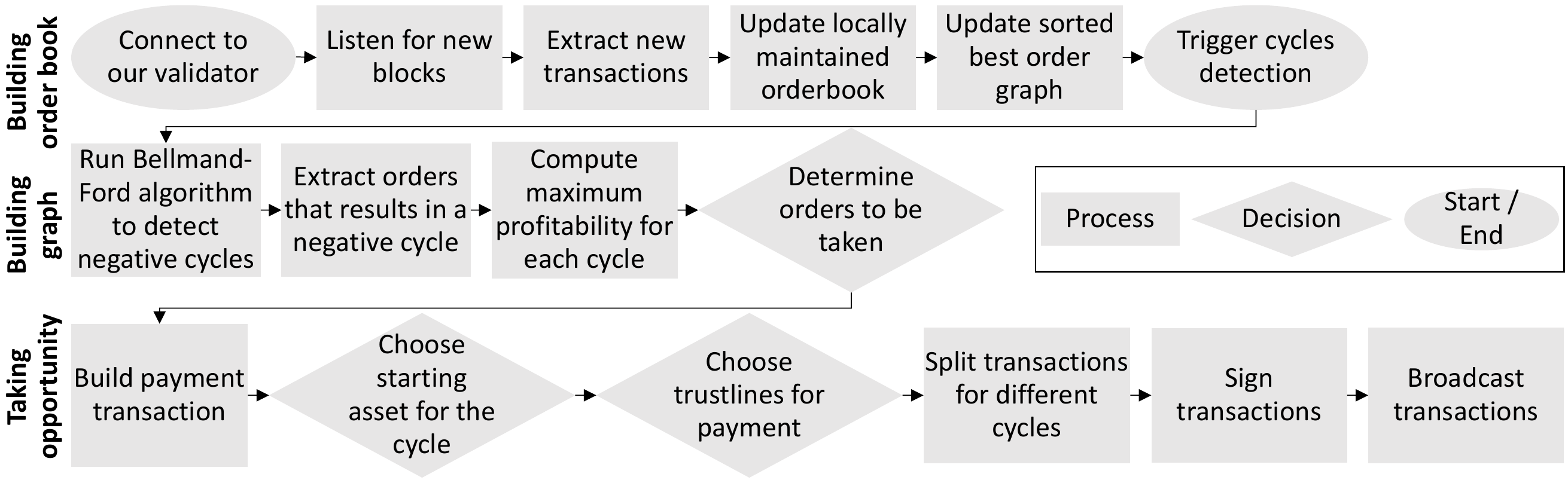}

\caption{{\sc Jack}'s arbitrage process.}
\label{fig:arbitrageproc}
\end{center}
\end{figure}

\autoref{fig:arbitrageproc} illustrates the arbitrage process employed by {\sc Jack}. Note that one round of arbitrage needs to be completed within one ledger interval, around 3.5 seconds.

\subsubsection{Building the order book}

We use the XRPL websocket APIs to listen to new blocks. 
At the launch of {\sc Jack}, the initial state of the order book is pulled with the {\tt book\_offers} request, which returns the current order book state of an issuer, and after the initial state is constructed, it is updated continually with every new block broadcast to the network. 

\subsubsection{Building the graph}

We use a modified Bellman-Ford algorithm \cite{Bellman1958} to build the graph.
We first sort orders upon their retrieval. We construct two lists of orders for each trading pair (in both directions). 
We then construct a graph with the ``best offer'' (highest bid, or lowest ask) for each pair as the only edge between the two currencies.
While new orders are added to the order book at each new block, we keep track of whether or not we modified the position of a previous ``best offer" in our sorted lists. If not, the graph remains the same with no need to re-run the cycle detection algorithm. If at least one edge has been modified due to the new block, we replace all the modified edges in the smaller graph and run cycle detection again. Thus, the algorithm ({\bf Algorithm~\ref{algo:bellman}}) progressively relaxes all edges on the graph, updating the shortest distance to each edge. 

\input{exhibits/bellman}

Running this algorithm $|v|-1$ times, we are able to obtain the shortest distance to each edge. Running one extra iteration on this graph, we can determine if there are negative cycles~\cite{Cherkassky1999Negative-cycleAlgorithms} in the graph, depending on whether or not the minimum distances were updated after this iteration. A negative cycle represents a path with a negative cost, i.e. a positive gain.

\subsubsection{Taking an opportunity}

Upon detection of a valid negative cycle, our bot \textsc{Jack} is able to build transaction paths according to the cycle. We then select the corresponding trustline to issue our payment and sign the transaction before publishing it through our validator.

\subsection{Caveats}

Risks associated with the usage of arbitrage bots still exist.

\subsubsection{Incomplete cycle}
We may identify opportunities of arbitrage that become invalid in the next ledger, due to pending transactions being fulfilled by other actors prior to our bot. Such unsuccessful attempts still incur transaction fees, resulting in a pure loss.

\subsubsection{Trusted actors}
\coin{XRP} is the only native asset in the ledger. Other currencies are issued in the form of IOU debt by issuing gateways who are responsible for honoring the debt~\cite{BecomePortal}. The solvency of those gateways cannot be guaranteed and there always exists a certain degree of irredeemability of IOU tokens. We have chosen for our experiment to open trustlines only with issuers featured on the list provided by Ripple to mitigate the counterparty risk.

\section{Conclusion}
We present the existing arbitrage behavior on XRPL DEX and introduce {\sc Jack the Rippler}, a bot that can automatically detect and take arbitrage opportunities. We discuss the limitations of our algorithm and argue that there still exist risks associated with using arbitrage bots.

%% file: exhibits/bellman.tex
\begin{algorithm}
\footnotesize
\KwResult{distance[], predecessor[]}
\KwIn{list $vertices$, list $edges$, vertex $source$}

// Initialization \\
\ForEach {vertex \ $v \in vertices$  \do}
{
distance[$v$] := inf \\
predecessor[$v$] := null 
}
distance[$source$] := 0

// Relax the edges repeatedly \\
\For{i \textbf{from} 1 \textbf{to} size(vertices)-1}
{\ForEach {$w$-weighted edge $(u, v) \in edges$ \do}
{
\If{distance[$u$] + $w$ < distance[$v$]}{
        distance[$v$] := distance[$u$] + $w$ \\
        predecessor[$v$] := $u$
   }
}
}

// Look for negative cycles \\
\ForEach {$w$-weighted edge $(u, v) \in edges$ \do}
{
\If{distance[$u$] + $w$ < distance[v]}{
        output "The graph contains a negative cycle"
   }
}

\Return distance[], predecessor[]

\caption{Bellman-Ford cycle detection}
\label{algo:bellman}
\end{algorithm}